\documentclass[aps,prd,twocolumn,amsmath,showpacs,superscriptaddress,nofootinbib]{revtex4}
\usepackage{graphicx}
\usepackage{longtable}
\usepackage{float}
\usepackage{dcolumn}
\usepackage{bm}
\usepackage{appendix}
\usepackage{multirow}
\usepackage{color}
\usepackage{soul}
\usepackage[normalem]{ulem}

\newcommand{\Mpc}{\mathrm{Mpc}}

\begin{document}

\title{Do joint CMB and HST data support a scale invariant spectrum? }

\author{Micol Benetti}

\author{Leila L. Graef}

\author{Jailson S. Alcaniz}

\affiliation{Departamento de Astronomia, Observat\'orio Nacional, 20921-400, Rio de Janeiro, RJ, Brasil}

\date{\today}

\begin{abstract}

{We combine current measurements of the local expansion rate, $H_0$, and Big Bang Nucleosynthesis (BBN) estimates of helium abundance with the latest cosmic microwave background (CMB) data from the Planck Collaboration to discuss the observational viability of the scale invariant Harrison-Zeldovch-Peebles (HZP) spectrum. We also analyze some of its extensions, namely, HZP + $Y_P$ and HZP + $N_{eff}$, where $Y_P$ is the primordial helium mass fraction and $N_{eff}$ is the 
effective number of relativistic degrees of freedom.}
We perform a Bayesian analysis and show that the latter model is favored with respect to the standard cosmology for values of $N_{eff}$ lying in the interval $3.70 \pm 0.13$  ($1\sigma$), which is currently allowed by some independent analyses. 
\end{abstract}

\pacs{98.80.Cq, 98.70.Vc, 98.80.Es}
\maketitle

{\emph{Introduction}} -- In the description of the early universe, scale-freeness was conjectured as a guiding cosmological principle over several decades. The scale invariant model with no tensor perturbation was proposed by Harrison \cite{Harrison:1969fb}, Zeldovich~\cite{Zeldovich:1972ij} and  Peebles~\cite{Peebles:1970ag}, long before realistic physical mechanisms of generation of density perturbations have been proposed. Such a spectrum, characterized by a spectral index $n_s = 1$, was proven in accordance with the early CMB data but became less attractive from the observational point of view as new and more precise data became available. The most recent result, using data from the second release of the Planck collaboration, shows that $n_s \neq 1$  at $5.6 \sigma$~\cite{Ade:2015lrj}\footnote{Using the same Planck data, the authors of Refs.~\cite{DiValentino:2015ola, DiValentino:2016hlg} showed that varying up to 12 cosmological parameters simultaneously, the value $n_s =1$ is off by $\sim 2.5$ standard deviations. For a different data set, Ref.~\cite{Zhang:2017epd} found that the scale-invariant spectrum is off by $\sim 3.3\sigma$.}. Theoretically, 
it is undeniable that the confirmation of this result, although not definitely proving the inflationary scenario \cite{Biswas:2013lna}, has important consequences  and points to the success of the theory of the quantum origin of cosmological perturbations and the early cosmic acceleration \cite{MUK,tpc}, which is the current paradigm for the early universe. 

However, the very same data used to obtain such a constraint on $n_s$ also provide, in the context of the $\Lambda$CDM model, a value of 
$H_0 = 67.31 \pm 0.96$ $\rm{km~s^{-1}~Mpc^{-1}}$  (68\% C.L.) for the local expansion rate {\cite{Ade:2015xua}\footnote{A lower value is obtained considering the new constraints on the reionization optical depth, obtained with Planck HFI data \cite{Aghanim:2016yuo}, i.e., $H_0 = 66.93 \pm 0.62$ $\rm{km~s^{-1}~Mpc^{-1}}$ at 68\% C.L.}. 
This value  differs by $\sim 3.4$ standard deviations from the value reported by Riess {\it et al.}~\cite{Riess}, 
$H_0 = 73.24 \pm 1.74$ $\rm{km~s^{-1}~Mpc^{-1}}$ (68\% C.L.), using four geometric distance calibrations of Cepheids. Since these two approaches sample different epochs in cosmic evolution, and one of them uses the $\Lambda$CDM model as the fiducial cosmology, this tension in the value of $H_0$ has given rise to speculations. It has been argued that either new physics beyond the standard cosmology or the influence of local structures  must be taken into account in order to reconcile these $H_0$ estimates~\cite{alternative1,alternative2} (see also \cite{others}). 

From CMB analyses, it is well known that $n_s$ and $H_0$ are positively correlated {(see, e.g., Fig. 1 of \cite{Gerbino:2016sgw})}. Therefore, given the arguments above, a legitimate question to ask is whether by combining the model-independent measurement of $H_0$~\cite{Riess} with current CMB data~\cite{Ade:2015lrj} the Harrison-Zeldovich-Peebles (HZP) spectrum and  some of its extensions are in fact ruled out as a viable observational description of the primordial spectrum. {In what follows we answer this question by performing, for the first time, a Bayesian analysis using the latest Planck data combined with the local $H_0$ measurement of Ref.~\cite{Riess} and estimates of the primordial helium abundance~\cite{Aver:2011bw,Peimbert:2007vm}. We compare the standard $\Lambda$CDM model, where the spectral index is allowed to vary, with the minimal HZP spectrum and two of its extensions, namely, HZP + $N_{eff}$ and HZP + $Y_P$.}  The 
effective number of relativistic degrees of freedom, $N_{eff}$,  and the primordial helium abundance, $Y_P$, are of particular interest as they are almost degenerated with $n_s$, both altering the damping tail of the temperature spectrum and mimicking a spectral index~\cite{Ade:2015xua, Hou:2011ec, Trotta:2003xg}. {{It is worth mentioning that $N_{eff}$ is not necessarily associated with new neutrino physics but refers to  any non-standard energy-density that took place in the early universe (see, e.g., \cite{ref1,ref2}).}}

%
\begin{table}[!]
    \centering
    \begin{tabular}{|c|c|}
        \hline
        Parameter   & Prior \\
        \hline  
        \hline
		$100\,\Omega_b h^2$ 	
		&[ $ 0.005 : 0.1 $ ]
		\\
		$\Omega_{c} h^2$	
		&[ $ 0.001 : 0.99 $ ]
		\\
		$100\, \theta$ 
		&[ $ 0.5 : 10 $ ]
		\\
		$\tau$
		&[ $ 0.01 : 0.8 $ ]
		\\
		$\ln 10^{10}A_s$\footnotemark[1]
		\footnotetext[1]{$k_0 = 0.05\,\Mpc^{-1}$.}
		&[ $ 2.0 : 4.0 $ ]
		\\
		$Y_P$ 
		&[ $ 0.1 : 0.6 $ ]
		\\
		$N_{eff}$ 
		&[ $ 2 : 5 $ ]\\
		\hline  
    \end{tabular}
    \caption{\label{tab:priors} Priors on the model parameters}
\end{table}
%
\begin{table*}[]
\centering
\caption{{
$68\%$ confidence limits for the cosmological parameters  using PLC data. 
The $\Delta \chi^2_{best}$ and the $\ln {B}_{ij}$ refers to the difference with respect to the $\Lambda$CDM.}
\label{tab:Tabel_results_1}}
\begin{tabular}{c|c|c|c|c}
\hline
{Parameter}&
{\textbf{$\Lambda$CDM}}& 
{\textbf{HZP}}& 
{\textbf{HZP+$Y_P$}}&
{\textbf{HZP+$N_{eff}$}}
\\
\hline
$100\,\Omega_b h^2$ 	
& $2.222 \pm 0.022$ 
& $2.300 \pm 0.020$ 
& $2.301 \pm 0.019$ 
& $2.294 \pm 0.019$ 

\\
$\Omega_{c} h^2$	
& $0.1197 \pm 0.0021$ 
& $0.1099 \pm 0.0011$ %
& $0.1151 \pm 0.0016$ %

\\
$100\, \theta$ 
& $1.04085 \pm 0.00045$ 
& $1.04217 \pm 0.00041$ 
& $1.04317 \pm 0.00045$ 
& $1.04054 \pm 0.00048$ 

\\
$\tau$
& $0.077 \pm 0.018$
& $0.140 \pm 0.017$
& $0.115 \pm 0.017$
& $0.109 \pm 0.017$	
\\
$\ln 10^{10}A_s$  \footnotemark[1]
\footnotetext[1]{$k_0 = 0.05\,\Mpc^{-1}$.}
& $3.088 \pm 0.034$ 
& $3.189 \pm 0.034$ 
& $3.166 \pm 0.034$ 
& $3.165 \pm 0.034$ 
\\
$Y_P$ 
& $0.2466 \pm 0.0001$ \footnotemark[2] \footnotetext[2]{Derived parameter obtained from BBN consistency.}
& $0.2470 \pm 0.0001$ \footnotemark[2]
& $ 0.2965 \pm 0.0098$ 
& $0.2553 \pm 0.0015$ \footnotemark[2]

\\
$N_{eff}$
& fixed to $3.046$ 
& fixed to $3.046$
& fixed to $3.046$ 
& $ 3.69 \pm 0.12$ 
\\
$H_0$
& $ 67.32 \pm 0.95 $ 
& $ 72.03 \pm 0.51 $
& $ 70.38 \pm 0.60 $ 
& $ 73.56 \pm 0.64 $ 
\\
\hline
\hline
$\Delta \chi^2_{\rm best}$         
& $ - $		
& $ 25.2$   
& $ 5.1 $	
& $ 4.4 $ 	
\\
$\ln \mathit{B}_{ij}$ 
& $-$ 
& $ - 25.8$ 
& $ - 3.2 $
& $ - 2.8 $ 
\\
\hline
\end{tabular}
\end{table*} 
%
\begin{table*}
\centering
\caption{{
$68\%$ confidence limits for the cosmological parameters using PLC+HST data.
The $\Delta \chi^2_{best}$ and the $\ln {B}_{ij}$ refers to the difference with respect to the $\Lambda$CDM.}
\label{tab:Tabel_results_2}}
%
\begin{tabular}{c|c|c|c|c}
\hline
{Parameter}&
{\textbf{$\Lambda$CDM model}}& 
{\textbf{HZP}}& 
{\textbf{HZP+$Y_P$ }}&
{\textbf{HZP+$N_{eff}$}}

\\
\hline
$100\,\Omega_b h^2$ 	
& $2.245 \pm 0.022$ 
& $2.302 \pm 0.019$ 
& $2.306 \pm 0.019$ 
& $2.294 \pm 0.019$ 
\\
$\Omega_{c} h^2$	
& $0.1167 \pm 0.0019$ 
& $0.1097 \pm 0.0010$ 
& $0.1144 \pm 0.0015$ 
& $0.1250 \pm 0.0033$ 
\\
$100\, \theta$ 
& $1.04130 \pm 0.00044$ 
& $1.04220 \pm 0.00040$ 
& $1.04318 \pm 0.00046$ 
& $1.04051 \pm 0.00051$ 
\\
$\tau$
& $0.091 \pm 0.019$
& $0.140 \pm 0.016$
& $0.116 \pm 0.018$ 
& $0.109 \pm 0.018$	
\\
$\ln 10^{10}A_s$  \footnotemark[1]
\footnotetext[1]{$k_0 = 0.05\,\Mpc^{-1}$.}
& $3.109 \pm 0.036$ 
& $3.189 \pm 0.034$ 
& $3.167 \pm 0.034$ 
& $3.164 \pm 0.034$ 
\\
$Y_P$ 
& $0.2467 \pm 0.0001$ \footnotemark[2] \footnotetext[2]{Derived parameter obtained from BBN consistency.} 
& $0.2470 \pm 0.0001$ \footnotemark[2]
& $0.2940 \pm 0.0099$ 
& $0.2553 \pm 0.0015$ \footnotemark[2]
\\
$N_{eff}$ 	 
& fixed to $3.046$ 
& fixed to $3.046$ 
& fixed to $3.046$
& $ 3.70 \pm 0.13 $ 
\\
$H_0$
& $ 68.74 \pm 0.87 $ 
& $ 72.13 \pm 0.46 $
& $ 70.70 \pm 0.58 $ 
& $ 73.54 \pm 0.60 $ 
\\
\hline
\hline
$\Delta \chi^2_{\rm best}$         
& $ - $		
& $ 23.4 $  
& $ 5.4 $	
& $ 4.1 $	
\\
$\ln \mathit{B}_{ij}$ 
& $ - $		 	
& $ - 7.2 $ 	
& $ + 4.1 $ 	
& $ + 6.9 $ 	
\\
\hline
\end{tabular}
\end{table*} 
{\emph{Method}} -- 
In order to calculate the Bayesian evidence factor, we implement the nested sampling algorithm of {\sc MultiNest} code~\cite{Feroz:2008xx,Feroz:2007kg,Feroz:2013hea} in the more recent release of the package {\sc CosmoMC}~\cite{Lewis:2002ah}. In our  analysis we use the most accurate Importance Nested Sampling (INS)~\cite{Cameron:2013sm, Feroz:2013hea} instead of the vanilla Nested Sampling (NS), requiring a INS Global Log-Evidence error of $\leq 0.02$ .

We consider the minimal $\Lambda$CDM model as the reference model, with the usual set of cosmological parameters: the baryon density, $\Omega_bh^2$, the cold dark matter density, $\Omega_ch^2$, the ratio between the sound horizon and the angular diameter distance at decoupling, $\theta$, the optical depth, $\tau$, the primordial scalar amplitude, $A_s$, and the primordial spectral index $n_s$. For the HZP model we consider $n_s$ fixed to unity whereas for the models HZP+$N_{eff}$ and HZP+$Y_P$ we add as free parameters, to the minimal set up, $N_{eff}$ and $Y_P$, respectively. 
In our analysis we choose to work with flat priors, as shown in Tab~\ref{tab:priors}. We consider purely adiabatic initial conditions and fix the sum of neutrino masses to $0.06~eV$. When the effective number of relativistic degrees of freedom is not treated as free parameter, $N_{eff}$ is fixed to $3.046$\footnote{The standard $\Lambda$CDM model assumes three massless neutrino families, quantifying their effects through the effective number of relativistic species, $N_{eff} = 3.046$. Note that $N_{eff} \neq N_{\nu} = 3$, to account for the fact that neutrinos are not completely decoupled during electron-positron annihilation, among other effects (see, e.g., Refs. \cite{Lesgourgues:2006nd, Mangano:2005cc, deSalas:2016ztq}).}.
At the same time, when the primordial helium mass fraction is not treated as free parameter, its value is derived from the BBN consistency relation, e.g., using the PArthENoPE fitting table\footnote{PArthENoPE
website: http://parthenope.na.infn.it/} to calculate the primordial abundances of helium and deuterium as a function of baryon density and the extra relativistic number of species. Finally, in addition to the parameters above we also vary the nuisance foregrounds parameters~\cite{Aghanim:2015xee}.

We use the CMB data set from the latest release of the Planck Collaboration~\cite{Ade:2015xua}, considering the high-$\ell$ Planck temperature data (in the range of $30< \ell <2508$) from the 100-,143-, and 217-GHz half-mission T maps, and  the low-P data by the joint TT,EE,BB and TE likelihood (in the range of $2< \ell <29$). Hereafter we refer to this dataset as ``PLC''. We also use the Riess {\it{et al.}} (2016) results on the local expansion rate, $H_0 = 73.24 \pm 1.74$ $\rm{km.s^{-1}.Mpc^{-1}}$ (68\% C.L.), based on direct measurements made with the Hubble Space Telescope \cite{Riess}. This measurement is used  as an external Gaussian prior and we refer to this joint data set as ``PLC+HST''.

{The observational viability of the three models considered in our analysis is discussed and compared with the reference model ($\Lambda$CDM) by calculating their corresponding Bayesian evidence, ${\cal{E}}_i$, and Bayes factor,  $\mathit{B}_{ij} = {\cal{E}}_i/{\cal{E}}_j$, where ${\cal{E}}_j$ is the evidence of the reference model. The Bayesian model comparison} is a very powerful tool to reward the models that fit well the data exhibiting strong predictivity, while models with a large number of free parameters, not required by the data, are penalised for the wasted parameter space.
The usual scale employed to judge differences in
evidence from the models is the Jeffreys scale \cite{Jeffreys}, which gives empirically calibrated levels of significance for the strength of evidence. 
In this work we will use a revisited and more conservative version of the Jeffreys convention suggested in \cite{Trotta}. In this convention,  
$ \ln \mathit{B}_{ij}  = 0 - 1 $,
$ \ln \mathit{B}_{ij}  = 1 - 2.5 $,
$ \ln \mathit{B}_{ij}  = 2.5 - 5 $,
and $ \ln\mathit{B}_{ij}  > 5 $ indicate an {\textit{inconclusive}}, {\textit{weak}}, {\textit{moderate}} and {\textit{strong}} preference of the considered model {with ${\cal{E}}_i$} with respect to the reference model {with ${\cal{E}}_j$}.
Note that, for an experiment which provides $\ln \mathit{B_{ij}} < 0$, it means support in favour of the reference model (see ref.~\cite{Trotta, Santos:2016sti} for a more complete discussion about this scale).\\

{\emph{Results}} -- The main quantitative results of our analysis using both the PLC and PLC+HST data sets are shown in the Tables~\ref{tab:Tabel_results_1} and \ref{tab:Tabel_results_2}, respectively.
In the last lines of these tables, we show the $\Delta\chi^2$ value and the Bayes factor of each model considering as reference the minimal $\Lambda$CDM scenario, as described above.
When only the PLC data are used, the analysis shows \textit{strong} and \textit{moderate} preference for the $\Lambda$CDM model with respect to the HZP, HZP + $N_{eff}$ and HZP + $Y_P$, respectively.
Moreover, we observe (Tab. \ref{tab:Tabel_results_1}) that the HZ model and its extensions require significantly higher values for the baryon density, as well as for reionization optical depths and primordial scalar amplitude, when compared to $\Lambda$CDM model. This result is in full agreement with the first analysis of the Planck Collaboration~\cite{Planck:2013jfk}, and the $\Delta\chi^2$ values also confirm its more recent results~\cite{Ade:2015xua}. 
In particular, we also note that the introduction of the $N_{eff}$ parameter worsens the precision of the  measurement of $\Omega_{c} h^2$ (see also~\cite{Planck:2013jfk}).

Comparing the results of the standard cosmology for the PCL and PCL + HST data sets, we note that the $\Lambda$CDM evidence for the latter  is $\sim 9$ times smaller than this same quantity when the former data set is considered, whereas the difference in $\chi^2$ between the two analysis is $\sim 1$. 
On the other hand, the Bayesian evidence for the HZP model improves by a factor of 10 when the HST prior is considered, which explains the increase of the value of its Bayes factor, $\ln \mathit{B}_{ij} = -7.2$, shown in Table III, with respect to the value displayed in Table II. In spite of this improvement, however, the minimal HZP model is still \textit{strongly} disfavored with respect to the $\Lambda$CDM model. The best-fit and evidence values of the HZP extensions show no significant changes when the $H_0$ prior is added. However, due to the drastic reduction of the $\Lambda$CDM model evidence, the Bayes factor for the HZP + $Y_P$ and  HZP + $N_{eff}$ models now shows, respectively, a \textit{moderate} and \textit{strong} preference of these models with respect to the standard cosmology. 

An important aspect  worth emphasizing concerns the result for the HZP + $Y_P$ model. Note that, although providing a better description for the PCL + HST data than the standard $\Lambda$CDM model, such a result is obtained at the cost of increasing significantly the value of the primordial helium abundance, $Y_P = 0.2940 \pm 0.0099$. This value is more than $4\sigma$ off from the central values obtained from direct measurements, $Y_P = 0.2534 \pm 0.008$,~\cite{Aver:2011bw} and from the standard big bang nucleosynthesis (BBN) estimate, $Y_P = 0.2477 \pm 0.0029$~\cite{Peimbert:2007vm, bbn}. Therefore, we conclude that this particular extension of the HZP spectrum cannot be regarded as a viable description of the primordial spectrum.  
On the other hand, the same conclusion cannot yet be drawn for the HZP + $N_{eff}$ model, whose prediction considering the  PCL + HST data set is $N_{eff}=3.70 \pm 0.13$. Indeed, our result matches the Planck Collaboration 1-parameter $\Lambda$CDM extension analysis at $2\sigma$, considering PLC data~\cite{Ade:2015xua}. 
Although considerably far from the standard value, $N_{eff} = 3.046$, recent BBN and CMB data also allow for values of $N_{eff}>3$ (see, e.g.~\cite{Ade:2015xua, DiValentino:2016hlg, Nollett:2013pwa,Benetti:2013wla, Hinshaw:2012aka, Boehm:2012gr, Cheng:2013csa}), which hampers a definitive conclusion on the observational viability of this particular extension of the HZP spectrum.\\

{\emph{Conclusion}} -- The currently most precise measurement of the local expansion rate $H_0$~\cite{Riess} differs by $\sim 3.4$ standard deviations from the value reported by the Planck Collaboration~\cite{Ade:2015lrj} assuming the $\Lambda$CDM model. This is the so-called $H_0$ tension discussed by several authors (see, e.g., \cite{Bernal:2016gxb} for a comprehensive study). In this paper the effect of this discrepancy on the estimates of the primordial spectrum  has been explored through a Bayesian analysis with the Planck and $H_0$ data. We have discussed the observational viability of the HZP spectrum and two of its extensions, i.e., HZP$+Y_P$ and HZP$+N_{eff}$ models. We have found that the latter extension is strongly favored over the $\Lambda$CDM model ($\ln \mathit{B}_{ij} = 6.9$) if $N_{eff}$ lies in the range $3.70 \pm 0.13$. Such result provides a clear example reinforcing the need for more precise, accurate and model-independent measurements of the local expansion rate than currently available~\cite{Suyu:2012ax}.\\

{\emph{Acknowledgments}} --
MB acknowledges financial support of the Funda\c{c}\~{a}o Carlos Chagas Filho de Amparo \`{a} Pesquisa do Estado do Rio de Janeiro (FAPERJ - fellowship {\textit{Nota 10}}). 
LG is supported by Coordena\c{c}\~{a}o de Aperfei\c{c}oamento de Pessoal de N\'{i}vel Superior (CAPES) (88887.116715/2016-00). 
JSA is supported by Conselho Nacional de Desenvolvimento Cient\'{i}fico e Tecnol\'{o}gico (CNPq) and FAPERJ. We also acknowledge the authors of the CosmoMC (A. Lewis) and Multinest (F. Feroz) codes. 

\end{document}